\documentclass[12pt]{article}
\usepackage[dvips]{graphics}
\usepackage{epsfig,graphicx}
\usepackage{amsfonts}
\usepackage{amsmath}

\input xy
\xyoption{all}

\begin{document}

\vskip 40 pt

\noindent {\large \bf Generalized q-deformed oscillators,
$q$-Hermite polynomials, generalized coherent states} \vskip 10 pt

{I.M. Burban}

\noindent Institute for Theoretical Physics, Ukrainian National
Academy of Sciences, Metrologichna st. 14b, Kyiv 03143, Ukraine

\vskip  40 pt


\vskip 40 pt

\noindent{\bf Abstract.} The aim of this paper is to study
generalized $q$-analogs of the well-known q-deformed harmonic
oscillators and to connect them with q-Hermite polynomials. We
give a construction of the appropriate oscillator-like algebras
and show that corresponding Hermite polynomials are generalization
of the discrete q-Hermite I and the discrete q-Hermite II
polynomials. We also construct generalized coherent states of
Barut-Girardello type for oscillator-like systems connected with
these polynomials.

\vskip 35 pt

\noindent {\bf 1. Introduction}
\medskip

\noindent

The simplest deformation of the canonical commutation relations
has been emerged in context of the study of the dual resonance
models of the strong interaction theories \cite {AC}. More general
deformaion of these relations was considered in connection with
description of representations of quantum groups ~\cite{B},
\cite{Mac}. It was introduced in order to extend the method of
realization of generators of Lie algebras by creation and
annihilation operators (the Jordan-Wigner construction) to the
quantum case. Since then various generalized q-analogs of these
deformations became the subject of mathematical and theoretical
physics. Their relation to the noncommutative geometry, special
functions of the q-analysis and other subjects of the mathematics
became evident. A connection between harmonic oscillators and
Hermite polynomials in the quantum mechanics is well-known. This
connection was generalized to the $q$-deformed cases as well. The
generalized $q$-deformed oscillators \cite{DK}, \cite{DK1},
\cite{BD}, \cite {BDY} are related to the $q$-deformed Hermite
polynomials in the same way as the standard quantum oscillator is
connected with the classical Hermite polynomials. In the work
\cite {BK3} the spectra of the position $Q$ and the momentum $P$
operators for various $q$-deformed oscillators in the Fock
representation has been described. There  the spectral measures
and the generalized eigenfunctions of these operators has been
found. They are expressed in terms of certain $q$-Hermitian
polynomials. Various orthogonal deformed $q$-Hermite polynomials
can appear depending on the type of a deformation of the
oscillator algebra.

 For example, in the case of Arik-Coon deformation with parameter
$q > 1$ one gets Hermite polynomials for which orthogonal measure
is known. The same problems has been studied for more general form
of the operators  $Q$ and the $P$  \cite {C-WK}. In this case the
spectral measure and the eigenfunctions has been expressed in
terms of the discrete $q$-Hermite polynomials. The spectral
measure of the position operator of Biedenharn-McFarlane
oscillator has been calculated in the case indetermine Hamburger
moment problem \cite {BDK}. Coherent states of the oscillator-like
systems, connected with some q-Hermite polynomials have been
constructed in \cite{BD}, \cite {BD1}.

Naturally, a problem of generalization of these results to the
case of other $q$-deformations of the harmonic oscillator algebra
arises. In this paper we consider oscillator-like systems giving a
description of the generalized $q$-deformed oscillators and
connect them with  the generalized discrete $q$-Hermite I and
q-Hermite II polynomials. These systems involve, as particular
cases, the known one-parameter deformations of oscillator
algebras.

In Section 2, we give a structure function, defining relations and
the position and momentum  operators $Q$ and $P$ of the
corresponding deformed oscillator algebra. In the Fock
representation the operators $Q$ and $P$ have a Jacobi matrix
form. We investigate the self-adjointness properties of these
operators.

In Section 3 we give examples of the oscillator-like systems
connected with discrete q-Hermite I and the generalized discrete
q-Hermite I polynomials. We establish the orthogonality relations
for these polynomials and, as a consequence, obtain a spectrum of
the position operators of these systems.

The same problem is solved in Section 4 for discrete Hermite II
and generalized discrete Hermite II polynomials of the
corresponding oscillator algebras.

In Section 5 we study generalized coherent states of oscillator
algebras corresponding to discrete q-Hermite I, generalize
discrete q-Hermite I and II polynomials on the basis of the method
suggested in \cite{BD}, \cite {BD1}.

The generalized $q$-deformed oscillator and its Heisenberg-Weyl
algebra is defined by the structure function $f(n)= f_n$ (a
positive function satisfying $f(0) = 0 $), which fix an
associative algebra generated by the elements $\{1, a, a^+, N \},$
satisfying the defining relations
\begin{equation}
\label{burban:cond0}
[N, a]= - a,\quad [N, a^+]= a^+,
\end{equation}
\begin{equation}
\label{burban:cond1} a^+a = f(N),\quad  aa^+ =
f(N+1).\end{equation} The structure functions $f(n)$ characterize
the deformation scheme. The Fock realization of these relations
and the number of particles operator have the form
\begin{equation}
\label{burban:repr1}
a|n\rangle = f_{n-1}|n-1\rangle,\quad a^+
|n\rangle = f_{n}|n+1\rangle,
\end{equation}

\begin{equation}\label{burban:repr2} N|n\rangle = n|n\rangle.
\end{equation}

As an example we consider the special case $f_{n}= (n+1)^{1/2}$ to
get the Heisenberg-Weyl algebra of the quantum oscillator of
quantum mechanics. It is generated by the generators $a, a^+, N$
and its defining relations are
\begin{equation}
[N, a]= - a, \quad [N, a^+]= a^+,\end{equation}
\begin{equation}
\label{burban:relation} [a, a^+] = 1,\quad [a, a] = [a^+, a^+] =
0.
\end{equation}
Recall that the Fock realization of this Heisenberg-Weyl algebra
is
\begin{equation}
\label{burban:represent} a|n\rangle = \sqrt{n}|n-1\rangle,\quad
a^+ |n\rangle = \sqrt{n+1}|n+1\rangle,\end{equation}
\begin{equation} N|n\rangle = a^+a|n\rangle = n |n\rangle.
\end{equation}
The position $Q$ and the momentum $P$ operators are unbounded
operators defined on a dense domain in the Hilbert space ${\cal
H}$ and satisfy the famous commutation relation $[Q, P]= i I.$
These operators are related to the creation and annihilation
operators $a^+$ and $a$ from (\ref{burban:represent}) by the
formulae
\begin{equation} Q = a^+ + a,\quad P = \frac {1}{i}(a^+ - a).
\end{equation}
Each of these operators have a Jacobi matrix form and due to the
condition
\begin{equation}
\sum_{n=0}^{\infty}\frac{1}{\sqrt{n}}= \infty
\end{equation}  is a self-adjoint
operator (Carleman's Lemma) \cite {Ber}. There is a well-known the
connection between the quantum-mechanical position  and momentum
operators $Q$ and $P$ and the Hermite polynomials $H_n(x).$ The
spectrum of the operator $Q$ is continuous and in order to find
its generalized eigenvectors we have to describe solutions of the
equation $Q |x\rangle = x |x\rangle.$ To do this, ones consider
the expansion $|x\rangle = \sum_{n=0}^{\infty}P_n(x)|n\rangle$ of
the vector $|x\rangle$ in the Fock space ${\cal H}.$ Using the
equations (\ref{burban:represent}) and $Q |x\rangle = x |x\rangle$
we obtain the following recurrence relation
\begin{equation} (n+1)^{1/2}P_{n+1}(x) + n^{1/2} P_{n-1}(x) = x
P_n(x),\quad n = 0,1,\ldots\end{equation} for coefficients
$P_n(x).$ This relation has the solutions
\begin{equation} P_n(x)=\frac{1}{(2^n\pi n!)^{1/2}} H_n(x),
\end{equation} with the initial conditions $ P_{-1}(x) = 0,\quad
P_0(x)= 1. $ The polynomials $H_n(x)$ satisfy the equation
\begin{equation} x H_n (x)= \frac{1}{2} H_{n +1}(x) + n
H_{n-1}(x) \end{equation} and can be written as
\begin{equation} H_n(x) =
\sum_{k=0}^{[n/2]}\frac{n!(-1)^k}{k!(n-2k)!}x^{n-2k}.
\end{equation}
They can be represented by means of hypergeometric function as
\begin{equation}
H_n(x)=(2 x)^n{}_2 F_0\Bigl(
\begin{array}{cc|c}
-n/2,&-(n-1)/2&\\{}&{-}&\end{array}-\frac{1}{x^2}\Bigr).
\end{equation}
The polynomials \begin{equation}\psi_n(x)= \Bigl(
\frac{1}{\sqrt{\pi }2^n n!}\Bigr )^{1/2}H_n(x)\end{equation} are
orthonormal with respect to the measure $ d \omega(x) = e^{-x^2}d
x$ and give wave functions of the Hamiltonian $H = a^+a + a a^+$
of the harmonic oscillator corresponding to the eigenvalues
$\lambda_n = 2 n + 1, n \ge 0.$

\medskip
\medskip
\medskip
\medskip
\noindent {\bf 2. Generalized oscillator algebras, position and
momentum operators and spectrum of Hamiltonian}
\medskip
\medskip
\medskip
\medskip

We consider oscillator-like systems defined by the structure
function \begin{equation} \label{burban:general} f_n = q^{\alpha
{(n+1)} + \beta/2}\Bigl(\frac {1-
q^{(l-1)(n+1)}}{1-q^{l-1}}\Bigr)^{1/2}
\end{equation} of oscillator algebra (see (Sec. 3) and (Sec. 4)). For the special values of the
parameters $\alpha, \beta,l,q$ this function reproduce known
versions of deformations of the oscillator Heisenberg-Weyl
algebra: the Arik-Coon, the Biedenharn-Macfarlane and the other
ones. The defining relations for these algebras can be written as
\begin{equation}
\label{burban:grelation1} a a^+ - q^{2\alpha} a^+ a =
q^{2\alpha(N+1)+ \beta}q^{(l-1)N},\end{equation}
\begin{equation}
\label{burban:grelation2}
aa^+ - q^{2\alpha + l-1}a^+ a =
q^{2\alpha(N + 1)+ \beta},
\end{equation}
or in a more compact form
\begin{equation}
\label{burban:grelation1}
aa^+ - q^{2\alpha} a^+ a =
q^{2\alpha(N+1)+ \beta}q'^N,
\end{equation}
\begin{equation}
\label{burban:grelation2} aa^+ - q^{2\alpha} q'a^ + a = q^{2\alpha
(N + 1) + \beta},
\end{equation}
where $q' = q^{l-1}.$ The Fock realization of the operators $a,$
$a^+$ in the Fock space ${\cal H}$ is
\begin{equation}
\label{burban:generalization1} a|n\rangle = q^{\alpha n +
\beta/2}\Bigl( \frac {1-q^{(l-1)n}}{1 - q^{l-1}}
\Bigr)^{1/2}|n-1\rangle,
\end {equation}
\begin{equation}
\label{burban:generalization2} a^+|n\rangle = q^{\alpha (n+1)
+\beta/2}\Bigl( \frac {1-q^{(l-1)(n+1)}}{1-q^{l-1}}
\Bigr)^{1/2}|n+1\rangle.
\end{equation}
A spectrum of the Hamiltonian $H = a^+a + aa^+$ of these
oscillator-like systems is discrete and is given by the expression
\begin{equation}\lambda_n = q^{2\alpha n + \beta}(1-q^{l-1})^{-1}\Bigl(
{1-q^{(l-1)n}} \Bigr) + q^{2\alpha (n+1) +
\beta}(1-q^{l-1})^{-1}\Bigl( {1-q^{(l-1)(n+1)}} \Bigr),
\end{equation} where $n\ge 0.$ However, wave functions
corresponding to these eigenvalues are defined in a more
complicated way. In the orthonormal basis $|n\rangle, n = 1, 2,
\ldots$ of the Hilbert space ${\cal H}$ the position and momentum
operators $Q$ and $P$ of these oscillators systems are given by by
the Jacobi matrices
\begin{equation}
\label{burban:gposition} Q |n\rangle = f_n |n+1\rangle +
f_{n-1}|n-1\rangle,\end{equation}\begin{equation} P|n\rangle =
\frac{1}{i}(f_n|n+1\rangle - f_{n-1}|n-1\rangle).\end{equation}
Unlike to the case of standard quantum oscillator, in this one the
self-adjointness of the operator $Q$ depends on the values of
parameters $q,\alpha, l.$ Due to Theorem 1.1, Chapter VII in
~\cite{Ber} the deficiency indices of the operator $Q$ are $(0,0)$
and then its closure ${\bar Q}$ is a self-adjoint operator, or
they are $(1,1)$ and then the operator $\bar{Q}$ allows
self-adjoint extensions. According to Theorem 1.5, Chapter VII in
~\cite{Ber}, if the function $f(n)$ from (\ref{burban:general})
satisfy the conditions
\begin {equation}
\label{burban:selfad} \quad f_{n-1}f_{n+1}\le f_n^2,\quad
\sum_{n=0}^{\infty}\frac{1}{f_n} < \infty,
\end{equation} then the deficiency indices of $Q$ are $(1,1).$
The first condition of (\ref{burban:selfad}) is reduced to the
inequality $q^{-(l-1)}- q^{l-1}\ge 2$ what is satisfied for all
positive $q.$ The convergence of the series (\ref{burban:selfad})
depends on the values of parameters of the deformation. Namely,
\begin{equation}
\label{burban:cas1} q < 1,\quad  \begin {cases} \alpha < 0,\quad
l-1 > 0,& \text{convergent, }\\ \alpha > 0,\quad l-1>0, &\text{
divergent, }\\ \alpha +l-1 < 0,\quad l-1 < 0,&\text{convergent,}\\
\alpha + l-1 > 0,\quad l-1< 0, & \text{divergent,}
\end{cases}
\end{equation}
and
\begin{equation}
\label{burban:cas2} q > 1,\quad  \begin {cases} \alpha < 0,\quad
l-1 >0,& \text{divergent,}\\ \alpha > 0,\quad l-1 >0, &\text{
convergent, }\\ \alpha +l-1 < 0,\quad l-1 < 0,
&\text{divergent,}\\ \alpha + l-1 > 0,\quad l-1 < 0, &
\text{convergent.}
\end{cases}
\end{equation}
In particular, this choice of the structure function unify the
following cases of the q-deformed of the oscillator algebras: the
Biedenharn-Macfarlane deformation ($ \alpha = 1/2,\quad \beta =
-1,\quad l = - 1,\quad q < 1 $), and ($ \alpha = - 1/2,\quad \beta
= 1,\quad l-1 = 2,\quad q > 1$)
\begin{equation}
\label{burban:bmrealization1} [N, a]= - a, \quad [N, a^+]= a^+,
\end{equation}
\begin{equation}
\label{burban:bmrealization2} aa^+ - q  a^+ a = q^{- N},\quad aa^+
- q^{-1}a^+ a = q^{N},
\end{equation}
and its symmetric generalization ($\alpha = 1/2,\quad \beta =
-1,\quad  l\in \mathbb{R}),$
\begin{equation}
\label{burban:srealization1} [N, a]= - a, \quad [N, a^+]= a^+,
\end{equation}
\begin{equation}
\label{burban:srealization2} aa^+ - q  a^+ a = q^{l N},\quad aa^+
- q^{l} a^+ a = q^{N},
\end{equation}
the deformation associated with the discrete $q$-Hermite I
polynomials ($\alpha = 1/2,\quad \beta = -1,\quad l = 2,\quad q <
1 $)
\begin{equation}
\label{burban:Irealization1}
[N, a]= - a, \quad [N, a^+]= a^+,
\end{equation}
\begin{equation}
\label{burban:Irealization2}
 aa^+ - q a^+ a = q^{2N},\quad aa^+ - q^2 a^+ a = q^N
\end{equation}
and the deformation associated with the discrete $q$-Hermite II
polynomials ($\alpha = -1,\quad \beta = 2,\quad l = 2,\quad q <
1$)
\begin{equation}
\label{burban:IIrealization1} [N, a]= - a,\quad [N, a^+]= a^+,
\end{equation}
\begin{equation}
\label{burban:IIrealization2} aa^+ - q^{-1} a^+ a = q^{-2 N},\quad
aa^+ - q^{-2}a^+ a = q^{-N}.
\end{equation}
\medskip

\medskip
\noindent {\bf 3. Generalized $q$-deformed oscillator-like systems
and generalized discrete q-Hermite I polynomials}

\medskip
\medskip
\noindent

First of all we consider a deformed oscillator in the case when
the parameters in ~(\ref{burban:general}) take the values $\alpha
= 1/2,\quad \beta = - 1,\quad l = 2,\quad 0 < q < 1.$ The
structure function ~(\ref{burban:general}) in this case is written
as
\begin{equation} f_n =
q^{(n+1)/2-1/2}(1-q)^{-1/2}(1-q^{n+1})^{1/2}.\end {equation} The
Fock representation of the $a$ and $a^+$ operators for
(\ref{burban:Irealization1}), (\ref{burban:Irealization2}) are
\begin{equation} a|n\rangle =
(\frac {1}{1-q})^{1/2}q^{(n-1)/2}(1-q^n)^{1/2}|n-1\rangle,
\end{equation} \begin{equation} a^+|n\rangle =
(\frac {1}{1-q})^{1/2}q^{n/2}(1-q^{n+1})^{1/2}|n+1\rangle.
\end{equation}
The deformed canonical commutation relations take the form
(\ref{burban:Irealization1}), (\ref{burban:Irealization2}) or
\begin{equation} [a, a^+] = q^N \frac{1-q^{N+1}}{1-q}-
q^{N-1}\frac{1-q^N}{1-q}.
\end{equation}
Recall that the Hamiltonian of this oscillator $H = aa^+ + a^+a$
has a discrete spectrum  $ H|n\rangle = \lambda_{n} |n\rangle,$
where \begin{equation} \lambda_n = q^{n}(1-q)^{-1}(1 - q^{n+1}) +
q^{n-1}(1-q)^{-1}(1-q^n), n \ge 0 .\end{equation} To find the wave
functions corresponding to these eigenvalues we proceed as in the
case of the standard quantum oscillator (see Introduction).

The position operator $Q$ and the momentum operator $P$ are given
in the basis $|n\rangle, n=0,1,\ldots$ of the Hilbert space $\cal
H$ by  Jacobi matrices.  The generalized eigenvectors
$\{|x\rangle\}$ of the operator $Q$, $Q|x\rangle = x |x\rangle,$
form a continuous basis of the Hilbert space ${\cal H}$ and
coefficients $P^{(0)}_n(x;q)$ of the transition from the basis
$\{|n\rangle \}$ to the basis $\{|x\rangle\},$ $ |x\rangle =
\sum_{n = 0}^{\infty} P^{(0)}_n (x;q)|n\rangle,$ satisfy the
recurrence relation $$ x P^{(0)}_n(x;q) = \Bigl(
\frac{1}{1-q}\Bigr)^{1/2}q^{n/2} $$
\begin{equation}
\label{burban:eq1}\times (1-q^{n+1})^{1/2}P^{(0)}_{n+1}(x;q) +
\Bigl(\frac{1}{1-q}\Bigr)^{1/2} q^{(n-1)/2}(1-q^n)^{1/2}P^{(0)}_
{n-1}(x;q).
\end{equation}
If we do the rescaling of variables  $y =(1-q)^{1/2}x$ and denote
$\psi^{(0)}_n(x;q)= P^{(0)}_n((1-q)^{-1/2}x;q),$ then the previous
relation is reduced to $$ x\psi^{(0)}_n(x;q)$$
\begin{equation} = q^{n/2
}(1-q^{n+1})^{1/2}\psi^{(0)}_{n+1}(x;q) +
q^{(n-1)/2}(1-q^n)^{1/2}\psi^{(0)}_{n-1}(x;q).
\end{equation}
After replacement \begin{equation} \label{burban:basis0}
\psi^{(0)}_n(x;q) = \frac {q^{-
n(n-1)/4}}{(q;q)_n^{1/2}}h^{(0)}_n(x;q),\end{equation} we obtain
the recurrence relation for the discrete $q$-Hermite I polynomials
\cite{GR}
\begin{equation}
\label{burban:hermitI} x\,{h}^{(0)}_n{(x;q)} =h^{(0)}_{n+1}(x;q) +
q^{n-1}(1-q^n){h}^{(0)}_{n-1}(x;q).
\end{equation} Together with the initial condition $h^{(0)}_0(x;q)= 1$ it
defines the discrete $q$-Hermitian I polynomials  \cite{GR},
\cite{KS} represented as
\begin{equation}
\label{burban:hermit} h^{(0)}_n(x;q) = \sum_{k=0}^{[n/2]}
\frac{(q;q)_n}{(q^2;q^2)_k (q;q)_{n-2k} }(-1)^k q^{k(k-1)}
x^{n-2k}.
\end{equation}
They can be written by means of the basic hypergeometric function
as
\begin{equation}
h^{(0)}_n(x;q)= x^n{}_2\phi_0\Bigl( \begin{array}{cc|c}
q^{-n},&q^{-n+1}&\\ {}&-&\end{array}q^2;\frac{q^{2n-1}}{x^2}
\Bigr).
\end{equation}
Now, the solution of the equation (\ref{burban:eq1}) can be
represented by the\quad expression
\begin{equation}
P^{(0)}_n(x;q) =
\frac{q^{-n(n-1)/4}}{(q;q)_n^{1/2}}h^{(0)}_n(\sqrt{1-q}x;q).
\end{equation}
It follows from ~\cite{GR},~\cite{KS} that these polynomials are
orthogonal with respect to the discrete measure $$
d\,\omega^{(0)}(x) = \frac{1}{2}(q;q^2)_{\infty}\,\delta (x -
\frac {q ^0}{\sqrt{1-q}})\,d x $$ $$+ \sum_{k > 0}
\frac{\sqrt{1-q}|x|}{2}\frac{(q^2(1-q)x^2, q;
q^2)_{\infty}}{(q;q)_{\infty}}\,\delta (x - \frac {q^k
}{\sqrt{1-q}})\,d x $$ \begin{equation}+ \sum_{k > 0}
\frac{\sqrt{1-q}|x|}{2}\frac{(q^2(1-q)x^2,q;
q^2)_{\infty}}{(q;q)_{\infty}}\,\delta(x +
\frac{q^k}{\sqrt{1-q}})\,d x.
\end{equation} and the orthogonality
relation is $$\frac{\delta_{mn}}{(q;q)_n} =
\frac{1}{2}(q;q^2)_{\infty}P^{(0)}_m(1;q)P^{(0)}_n(1;q) $$
\begin{equation}
+ \sum_{k = 0}^{\infty}\{P^{(0)}_m(q^k;q) P^{(0)}_n(q^k;q)+
P^{(0)}_m(-q^k;q)
P^{(0)}_n(-q^k;q)\}\frac{q^k}{2}\frac{(q^{2k+2},q
;q^2)_{\infty}(q;q^2)_{\infty}}{(q^2;q^2)_{\infty}}.
\end{equation}
It follows that spectrum of the position operator $Q$ is
\begin{equation}
Sp\,Q = \Bigl\{ \frac{\pm 1}{\sqrt{1-q}},\quad\frac{\pm
q}{\sqrt{1-q}}, \ldots,\frac{\pm q^k}{\sqrt{1-q}},\ldots; k \ge 0
\Bigr\}.
\end{equation}

The extension of this method for the generalized oscillator
(\ref{burban:general}), (\ref{burban:grelation1}),
(\ref{burban:grelation2}), determined by the formulas
(\ref{burban:generalization1}) and (\ref{burban:generalization2})
for $q < 1$ gives the recurrence relations

$$ x P_n{(x;q)} = \Bigl(\frac{1}{1-q'}\Bigr)^{1/2}q^{{\alpha (n +
1)} + \beta/2}(1-q'^{(n+1)})^{1/2}P_{n+1}(x;q) $$\begin{equation}
\label{burban:equl1} + \Bigl(\frac{1}{1-q')}\Bigr)^{1/2} q^{\alpha
n + \beta/2}(1-q'^{n})^{1/2}P_{n-1}(x;q).
\end{equation}
If we rescale the variables $y = (1-q')^{1/2} x, $ then $
P_n(x;q)= \psi_n ((1 - q')^{1/2}x;q)$ yields $$ x \psi_n(x;q)$$
\begin{equation}
\label{burban:equer} = q^{\alpha (n + 1) + \beta/2}(1 -
q'^{n})^{1/2}\psi_{n+1}(x;q) + q^{\alpha n + \beta/2}(1 -
q'^{n})^{1/2}\psi_{n-1}(x;q).
\end{equation}
Representing the function $\psi_n(x;q)$ as
\begin{equation}
\label{burban:basis1} \psi_n(x;q) = \frac {q^{-\alpha
n^2/2}}{q^{(\alpha + \beta)
n/2}(q';q')_n^{1/2}}h_n(x;q)\end{equation} we obtain from
(\ref{burban:equer}) the recurrent relation for the generalized
q-Hermite polynomials $h_n(x;q):$
\begin{equation}
\label{burban:discrhermit} x h_n(x;q)= h_{n+1}(x;q)+ q^{2\alpha n
+ \beta} (1-q'^{n})h_{n-1}(x;q).
\end{equation}
 This equation can be solved by means of the anzatz
\begin{equation}
\label{burban:ghermitI} h_n(x;q)= \sum_{k = 0}^{[n/2]}
\frac{(q';q')_n} {((a_n,c_n);(1,q'^d))_k (q';q')_{n-2k}}(-1)^k
q'^{k(k-1)}x^{n-2k},
\end{equation}
where we use the notation \cite{JR}
\begin{equation}
((a, c );(p,q))_k = \begin {cases} 1,& \text{if $ k = 0 ;$}\\ (a -
c)(a p - c q)\ldots(a p^{k-1}- c q^{k-1}),&\text{otherwise, }\\
\end{cases}
\end{equation} $q' = q^{l-1}$, and $a_n, c_n, d$ are unknown quantities.
It is easy to see that this anzatz leads the relation (\ref
{burban:discrhermit}) to the identity
\begin{equation}
\label{burban:identityI}
 1 - q'^{n+1}- q^{2\alpha n +
\beta}(a_n-c_n q'^{d(k-1)})q'^{-2(k-1)} = 1-q'^{n-2k+1}.
\end{equation}
This identity admits the solutions \begin{equation}
\label{burban:solutionI} a_n = q^{-2\alpha n -
\beta}q'^{n-1},\quad c_n = q^{-2\alpha n -\beta}q'^{n+1},\quad d =
2
\end{equation} and an easy calculation gives $((a_n,c_n);(1,q'^d))_k =
q^{-k(2\alpha n + \beta)}q'^{k(n-1)}(q'^2;q'^2)_k.$

The resulting expressions for the generalized q-Hermite
polynomials can be written as the polynomial of degree $n$ in $x$
\begin{equation}
\label{burban:finitaI}
 h_n(x;q)= \sum_{k = 0}^{[n/2]} \frac{(q';q')_n}{(q'^2;q'^2))_k
(q';q')_{n-2k}}(-1)^k q^{(2\alpha n + \beta)k} q'^{k(k-n)}
x^{n-2k}.
\end{equation}
They can be represented in terms of the basic hypergeometric
function as
\begin{equation}
h_n(x;q)= x^n{}_2\phi_0\Bigl( \begin{array}{cc|c}
q'^{-n},&q'^{-n+1}&\\ {}&-&\end{array}q'^2;\frac{q^{2\alpha n
+\beta} q'^n}{x^2} \Bigr).
\end{equation}

It is easy to see that for $\alpha = \frac{1}{2},\quad \beta =
-1,\quad l = 2$ the solution (\ref{burban:finitaI}) of
(\ref{burban:discrhermit}) reduces to the solution
(\ref{burban:hermit}) of (\ref{burban:hermitI}).

At last, the solutions $P_n(x;q)$ of the equations
(\ref{burban:equl1}) with the initial conditions $P_{-1}(x;q) =
0,\quad P_0(x;q) = 1$ can be written as polynomials of degree $n$
in $x:$
\begin{equation} \label{burban:resI} P_n(x;q) =
\frac{q^{-\alpha n^2/2}}{q^{\frac{\alpha + \beta}{2}
n}(q';q')_n^{1/2}}h_n (\sqrt{1-q'}x;q).
\end{equation}
\medskip
Now we restrict ourselves by the condition $\alpha = (l-1)/2$ in
(\ref{burban:resI}). Then \begin{equation} P_n(x;q)=
\frac{q^{-\alpha n(n-1)/2}}{(q';q')_n ^{1/2}}h_n^0 (q^{-(2\alpha +
\beta)/2} \sqrt{1-q'}x;q').
\end{equation}
These polynomials are orthogonal with respect to the discrete
measure $$ d\omega(x) = \frac {q^{-(2\alpha + \beta)}\sqrt{1-q'}
}{2}(q';q'^2)_{\infty}\delta (x - \frac {q'^0}{ q^{-(2\alpha +
\beta )}\sqrt{1-q'}})d x $$ $$+ \sum_{k > 0} \frac{q^{-(2\alpha +
\beta)/2}\sqrt{1-q'}|x|}{2}\frac{(q^{-(2 \alpha +
\beta)}(q'^2(1-q')x^2,q'; q^2)_{\infty}}{(q';q')_{\infty}}
\,\delta (x - \frac {q'^k}{q^{-(2 \alpha + \beta)/2}\sqrt{1-q'}})d
x $$
\begin{equation}
+ \sum_{k > 0} \frac{q^{-(2\alpha +
\beta)/2}\sqrt{1-q'}|x|}{2}\frac{(q^{-(2 \alpha +
\beta)}(q'^2(1-q')x^2,q'; q^2)_{\infty}}{(q';q')_{\infty}}
\,\delta ( x + \frac {q'^k}{q^{-( 2 \alpha +
\beta)/2}\sqrt{1-q'}})d x.
\end{equation}
The orthogonality relation has the form $$
\frac{\delta_{mn}}{(q';q')_n} =
\frac{1}{2}\frac{(q';q')_{\infty}}{(q'^2;q'^2)_{\infty}}
P_m(1;q)P_n(1;q)  $$
\begin{equation}
+ \sum_{k > 0}\{P_m (q'^k;q)P_n(q'^k;q) + P_m (- q'^k;q)P_n(-
q'^k;q)\}\frac{q'^k}{2}\frac{(q'^{2k+2},q';q'^2)_{\infty}(q';q'^2)_{\infty}}
{(q'^2;q'^2)_{\infty}}.\end {equation} From this it follows that
spectrum of the position operator $Q$ is
\begin{equation} Sp\,Q =
\Bigl\{ \frac{\pm q^{(2\alpha +
\beta)/2}}{\sqrt{1-q'}},\quad\frac{\pm q^{(2\alpha +
\beta)/2}q'}{\sqrt{1-q'}},\ldots, \frac{\pm q^{(2\alpha +
\beta)/2}q'^k}{\sqrt{1-q'}},\ldots; k\ge 0 \Bigr\}.
\end{equation}

\bigskip
\medskip
\medskip
\medskip

\noindent {\bf 4. Generalized $q$-deformed oscillator-like systems
and generalized discrete q-Hermite II polynomials}

\medskip
\medskip
\medskip

If we fix in (\ref{burban:general}) the values of the parameters
$\alpha = -1,\quad \beta = 2,\quad l= 2,\quad 0 < q < 1,$ the
structure function $f_n$ is reduced to the form
\begin{equation} f_n = q^{-(n+1)+1}(1-q)^{-1/2}(1-
q^{n+1})^{1/2}.\end{equation} The Fock representation of the
creation and the annihilation operators of the relations
(\ref{burban:IIrealization1}), (\ref{burban:IIrealization2}) is
given by
\begin {equation}
\label{burban:annihil} a|n\rangle =
\Bigl(\frac{q}{1-q}\Bigr)^{1/2}q^{-n+1/2}(1-q^n)^{1/2}|n-1\rangle,
$$ $$ a^+|n\rangle = \Bigl(\frac {q}{1-q}\Bigr)^{1/2}q^{-n-1/2}(1-
q^{n+1})^{1/2}|n+1\rangle.
\end{equation} It follows that
\begin{equation} aa^+|n\rangle =
q^{-2n}\frac{1-q^{n+1}}{1-q}|n\rangle,\quad a^+a|n\rangle =
q^{-2n+2}\frac{1- q^n}{1-q}|n\rangle,
\end{equation}
and commutation relation (\ref{burban:IIrealization2}) can be
written in the symbolic form
\begin{equation} [a, a^+] = q^{-2N}\,
\frac{1-q^{N+1}}{1 - q} - q^{-2 N+2}\, \frac{1-q^N}{1- q}.
\end{equation}
The Hamiltonian $H$ of this oscillator-like system has the
discrete spectrum $ H |n\rangle = \lambda_n |n\rangle,$ where
\begin{equation}\lambda_n =q^{-2n}(1-q)^{-1}(1- q^{n+1}) +q^{2-2n}(1-q)^{-1}
(1 - q^n), \quad n \ge 0.\end{equation} As in the previous section
the position and momentum operators $Q$ and $P$ in the basis
$|n\rangle$ of the Hilbert space ${\cal H}$ are represented by
Jacobi matrices. The coefficients ${\tilde P}^0_n(x;q)$ of the
transition $|x\rangle = \sum_{n = 0}^{\infty} {\tilde P}^0_n
(x;q)|x\rangle $ from the basis $\{|n\rangle\}$ to the basis
$\{|x\rangle \}$, , $Q|x\rangle = x|x\rangle,$ satisfy the
relations $$ x {\tilde P}^0_n(x;q) = \Bigl(
\frac{q}{1-q}\Bigr)^{1/2}q^{-(n +1) + 1/2} $$
\begin{equation}
\label{burban:rel2} \times(1-q^{n+1})^{1/2}{\tilde P}^0_{n+1}(x;q)
+ \Bigl(\frac{q}{1 - q}\Bigr)^{1/2} q^{-n+1/2}(1- q^n){\tilde
P}^0_{n-1}(x;q).
\end{equation}
Introducing the rescaling $y = q^{-1/2}(1-q)^{1/2} x$ and the
function $ {\tilde \psi}^{(0)}_n(x;q)= {\tilde
P}^{(0)}_n(q^{1/2}(1-q)^{-1/2}x;q)$ we obtain the equation $$
x{\tilde \psi}^{(0)}_n(x;q) = q^{-(n+1)+1/2}$$
\begin{equation}
\times(1- q^{n+1})^{1/2}{\tilde \psi}^{(0)}_{n+1}(x;q) +
q^{-n+1/2}(1 - q^n)^{1/2}{\tilde \psi}^{(0)}_{n-1}(x;q).
\end{equation}
After the replacement
\begin{equation}\label{burban:basis2}{\tilde\psi}^{(0)}_n(x;q)=
\frac{q^{n^2/2}}{(q;q)_n^{1/2}}{\tilde
h}^{(0)}_n(x;q),\end{equation} we obtain the recurrence relation
for the discrete $q$-Hermite II polynomials \cite{GR}
\begin{equation}
\label{burban:discrhermitII}x{\tilde h}^{(0)}_n{(x;q)} = \tilde
{h}^{(0)}_{n+1}(x;q) + q^{-2n+1}(1-q^n){\tilde
h}^{(0)}_{n-1}(x;q)\end{equation} which together with the initial
condition ${\tilde h}^{(0)}_0(x;q) = 1$ define the discrete
$q$-Hermite II polynomials
\begin{equation}\label{burban:hermitII} {\tilde h}^{(0)}_n(x;q) =
\sum_{k = 0}^{[n/2]} \frac{(q;q)_n}{(q^2;q^2)_k (q;q)_{n-2k}
}(-1)^k q^{2k(k-n)+k} x^{n-2k}.
\end{equation} These polynomials can be represented
in terms of the basic hypergeometric function:
\begin{equation} {\tilde h^{(0)}_n}(x;q)= x^n {}_2\phi_1\Bigl(\begin
{array}{c c|c} q^{-n}&q^{-n+1}&\\ {}&{0}&\end{array}
q^2;-\frac{q^2}{x^2}\Bigr).\end{equation} The solution of the
equations (\ref{burban:rel2}) with the initial conditions ${\tilde
P}^{(0)}_{-1}(x;q) = 0, {\tilde P}^{(0)}_0(x;q)= 1 $  can be given
in the form
\begin{equation} {\tilde P}^{(0)}_n(x;q) = \frac{q^{n^2/2}
}{(q;q)_n^{1/2}}{\tilde
h}_n^{(0)}(q^{-1/2}\sqrt{1-q}x;q).\end{equation} It follows from
~\cite{KS} that these polynomials are orthogonal with respect to
the discrete measure $$ d\,{\tilde \omega}^{(0)}(x)$$ $$ = \sum_{k
=-\infty}^{\infty}c^{-1}q^{-1/2}\sqrt{1-q}
w(q^{-1/2}\sqrt{1-q}x;q)\, x\, \delta (x -\frac {c q^k
}{q^{-1/2}\sqrt{1-q}})\,d x $$
\begin{equation}
-\sum_{k = -\infty}^{\infty} c^{-1}q^{-1/2}\sqrt{1-q}
w(-q^{-1/2}\sqrt{1- q}x;q)\,x\, \delta (x + \frac {c q^k
}{q^{-1/2}\sqrt{1-q}})\,d x.
\end{equation}
The orthogonality relation for these polynomials has the form
 $$ \sum_{k =
-\infty}^{\infty}\{{\tilde P}^{(0)}_m(cq^k;q) {\tilde
P}^{(0)}_n(cq^k;q) + {\tilde P}^{(0)}_m(-cq^k;q) {\tilde
P}^{(0)}_n(-cq^k;q)\}w(cq^k;q)q^k  $$
\begin{equation}
= 2\frac{(q^2,-c^2q,-c^{-2}q
;q^2)_{\infty}}{(q,-c^2,-c^{-2}q^2;q^2)_{\infty}}\,\delta_{mn},\quad
c > 0,\end{equation} where $ w(x;q) = 1/(-x^2;q^2)_{\infty}.$ It
follows that spectrum of the position operator $Q$ is
\begin{equation}
Sp\,Q =\Bigl\{ \frac{\pm c}{q^{-1/2}\sqrt{1-q}},\quad\frac{\pm c
q}{q^{-1/2}\sqrt{1-q}},\ldots,\frac{\pm cq^k}{q^{-1/2}\sqrt{1-q}},
\ldots; k\ge 0  \Bigr\}.
\end{equation}
A connection of the discrete q-Hermite I polynomials and the
discrete q-Hermite II polynomials are given by $q \to 1/q.$
Indeed, we have the relations
\begin{equation} \frac {(1/q';1/q')_n
q'^{-k(k-1)}}{(1/q';1/q')_{n-2k}}= \frac{(q';q')_n
}{(q';q')_{n-2k}} q'^{k^2-2kn}, \end{equation}
\begin{equation}
(1/q^2;1/q^2)_k = (-1)^k (q^2;q^2)_k q^{-k(k+1)}\end{equation}
leading to the identity \cite{KS}
\begin{equation}
\label{burban:change} h^{(0)}_n({\rm i}x;q^{-1}) = {\rm
i}^n\tilde{h}^{(0)}_n (x;q).
\end{equation}
This identity reflects the transition $q \to q^{-1}$ from the
oscillator (\ref{burban:Irealization1}),
(\ref{burban:Irealization2}) to the oscillator
(\ref{burban:IIrealization1}), (\ref{burban:IIrealization2}).

Now we consider the generalized oscillator (\ref{burban:general}),
(\ref{burban:grelation1}), (\ref{burban:grelation2}) represented
by operators (\ref{burban:generalization1}),
(\ref{burban:generalization2}), (\ref{burban:cas2}) for $q < 1.$
Then instead (\ref{burban:equl1}) we have the equality $$
x\,{\tilde P}_n(x;q) = \Bigl(\frac{q^{\beta/2}}{1-q'}\Bigr)^{1/2}
q^{\alpha(n+1)+ \beta/4} $$
\begin{equation}
\label{burban:resII}\times ( {1-q'^{n+1}})^{1/2}{\tilde
P}_{n+1}(x;q) + \Bigl(\frac{q^{\beta/2}}{1 - q'}\Bigr)^{1/2}
q^{\alpha n + \beta/4}(1 - q'^n)^{1/2}{\tilde P}_{n-1}(x;q),
\end{equation}
or $$x {\tilde \psi}_n  $$ \begin{equation} \label{burban:eher} =
q^{\alpha (n+1) + \beta/4}(1-q'^{n+1})^{1/2}{\tilde \psi}_{n +1
}(x;q) + q^{\alpha n + \beta /4} (1-q'^n )^{1/2}{\tilde
\psi}_{n-1}(x;q),\end{equation}  where ${\tilde \psi}_n (x;q) =
{\tilde P}_n((q^{\beta/4}(1-q'))^{-1/2}x;q).$ Representing the
function $ {\tilde \psi}_n(x;q)$ as
\begin{equation}
\label{burban:basis3} {\tilde \psi}_n(x;q) = \frac{q^{-\alpha
n^2/2}}{q^{(2\alpha + \beta)n/4}(q';q')_n^{1/2}}{\tilde h}_n(x;q)
\end{equation} we obtain the recurrence relation
\begin{equation}
\label{burban:dublrel} {\tilde h}_{n+1}(x;q)+ q^{2\alpha n +
\beta/2} (1-q'^{n}){\tilde h}_{n-1}(x;q) = x {\tilde h}_n(x;q).
\end{equation}
The solution of this equation can be obtained by means of the
anzatz
\begin{equation}
\label{burban:gdishermitII} {\tilde h}_n(x;q) = \sum_{k =
0}^{[n/2]} \frac{(q';q')_n}{((a_n,c_n);(1,q'^d))_k
(q';q')_{n-2k}}(-1)^k q'^{k (2k - 2n + 1)}x^{n-2k},
\end{equation}
which generalizes (\ref{burban:hermitII}). It is easy to see that
it reduces the relations (\ref{burban:dublrel}) to the identity $$
(1-q'^{n-2k+1})q'^{2k} $$
\begin{equation}
= 1 - q'^{n+1} - q^{2\alpha n + \beta /2}(a_n -c_n
q'^{d(k-1)})q^{-(2k-2n-1)}q'^{2(n-1)}.
\end{equation}
We obtain the solution
\begin{equation}
a_n = q^{-2\alpha n - \beta /2}q'^{-2n+1},\quad c_n = q^{-2\alpha
n-\beta/2}q'^{-2n+3},\quad d = 2.
\end{equation}
An easy calculation gives $((a_n,c_n);(1,q'^d))_k = q^{- k(2\alpha
n + \beta/2)}q'^{- k(2n-1)}(q'^2;q'^2)_k.$ The resulting
expression
\begin{equation}
\label{burban:finitaII} {\tilde h}_n(x;q)= \sum_{k = 0}^{[n/2]}
\frac{(q';q')_n}{(q'^2;q'^2)_k (q';q')_{n-2k}}(-1)^k q^{(2\alpha n
+ \beta/2)k}q'^{2k^2}x^{n-2k}
\end{equation}
defines a generalized of the q-Hermite polynomials which can be
written in terms of the basic hypergeometric function,
\begin{equation}
{\tilde h}_n(x;q)= x^n{}_2\phi_1\Bigl( \begin{array}{cc|c}
q'^{-n},&q'^{-n+1}&\\ {}& 0 &\end{array} q'^2;-\frac{q^{2\alpha n
+ \beta/2} q'^{2n+1}}{x^2} \Bigr).
\end{equation}

It is easy to see that for the special values  $\alpha = -1,\quad
\beta = 2,\quad l = 2$ the solution (\ref{burban:finitaII}) is
reduced to the solution (\ref{burban:hermitII}) of the relation
(\ref{burban:discrhermitII}).

Finally, the solution ${\tilde P}_n(x;q)$ with the initial
conditions ${\tilde P}_{-1}(x;q)= 0,\quad {\tilde P}_0(x;q)= 1$ of
the equation (\ref{burban:resII})  are given by the formula
\begin{equation}
\label{burban:eher1} {\tilde P}_{n}(x;q)= \frac{q^{-\alpha n^2/2}
} {{q^{(2 \alpha + \beta) n/4 }(q';q')_n^{1/2}}}{\tilde h}_n(q^
{-\beta/4}\sqrt{1-q'}x;q).
\end{equation}
From now on we restrict ourselves by the condition $\alpha = -
(l-1)$ in (\ref{burban:eher1}). Then
\begin{equation}
{\tilde P}_n (x;q) = \frac{q^{-\alpha n^2\,/2 }}{
(q';q')_n^{1/2}}{\tilde h}_n ^0 (q^{-(\alpha +\beta)
/2}\sqrt{1-q'}x;q').
\end{equation}
These polynomials are orthogonal with respect to the discrete
measure $$ d\,{\tilde \omega(x)}  $$ $$ =\sum_{k =
-\infty}^{\infty}c ^{-1}q^{-(\alpha+\beta)/2} \sqrt{1-q'}
w(q^{-(\alpha+\beta)/2}\sqrt{1-q'}x;q')\,x\, \delta (x - \frac {c
q'^k }{q^{-(\alpha+\beta)/2}\sqrt{1-q'}})\,d x $$
\begin{equation}
- \sum_{k = -\infty}^{\infty}c^{-1}q^{-(\alpha + \beta)/2}
\sqrt{1-q'} w(q^{-(\alpha + \beta)/2)}\sqrt{1-q'}x;q')\,x\, \delta
(x + \frac {c q'^k }{q^{-(\alpha+\beta)/2}\sqrt{1-q'}})\,d x
\end{equation}
and the orthogonality relation is $$ \sum_{k =
-\infty}^{\infty}\{{\tilde P}_m(cq'^k;q) {\tilde P}_n(cq'^k;q) +
{\tilde P}_m(-cq'^k;q) {\tilde P}_n(-cq'^k;q)\}w(cq'^k;q)q'^k  $$
\begin{equation}
= 2\frac{(q'^2,-c^2q',-c^{-2}q' ;q'^2)_{\infty}(q';q')_n
q^{-\alpha
n^2}}{(q',-c^2,-c^{-2}q'^2;q'^2)_{\infty}q'^{n^2}}\delta_{mn},\quad
c > 0,\end{equation} where $ w(x;q)= 1/(-x^2;q^2)_{\infty}.$ It
follows that spectrum of the position operator $Q$ is $$ Sp\,Q $$
\begin{equation}= \Bigl\{\frac{\pm c}{q^{-(\alpha
+\beta)/2}\sqrt{1-q'}},\quad\frac{\pm c q'}{q^{-(\alpha
+\beta)/2}\sqrt{1-q'}},\ldots,\frac{\pm cq'^k}{q^{-(\alpha
+\beta)/2}\sqrt{1-q'}},\ldots; k \ge 0 \Bigr\}.
\end{equation}

A connection of the generalized q-Hermitian I and the generalized
q-Hermitian II polynomials is not evident at all as in non
-generalized case. Unfortunately the change $q \to 1/q$ does not
lead to the analogous of (\ref{burban:change}) for the generalized
q-Hermitian I and the generalized q-Hermitian II polynomials
(\ref{burban:finitaI}) and (\ref{burban:finitaII}). It is evident
from (\ref{burban:generalization1}) and
(\ref{burban:generalization2}). Instead of this in this case we
have
\begin{equation}\label{burban:gItogII} h_n(x;1/q) =
\sum_{k=0}^{[n/2]} \frac{(q';q')_n }{(q'^2;q'^2)_k
(q';q')_{n-2k}}(-1)^k q^{-k(2\alpha n + \beta)}
q'^{k(2k-n)}x^{n-2k}.
\end{equation}
It is easy to see that for $\alpha = \frac{1}{2},\quad \beta = -
1,\quad l-1 = 1$ this relation gives identity (\ref
{burban:change}).

\medskip
\medskip
\medskip
\medskip

\noindent {\bf 5. Barut -Girardello coherent states of oscillators
associated with generalized discrete q-Hermite I and II
polynomials}
\medskip
\medskip
\medskip
\medskip

It is known that the coherent states in the ordinary Lie algebras
are very useful for studying the representation theory. The
generalized coherent states are very useful in the study of
representation of quantum group and physics, in particular, in
quantum optics. Barut-Girardello type coherent states of
oscillator algebras have been studied for oscillator-like system
connected with some orthogonal polynomials. The family coherent
states associated with discrete q-Hermite polynomials of type II
have been described in ~\cite {BD} and ~\cite{BD1}. In this
section we give solution the same problem for discrete q-Hermite I
and generalized discrete q-Hermite I and II polynomials.

First of all we prove the formula for a generating function of
q-Hermite I polynomials connected with the appropriate
q-oscillator
\begin {equation}
\label{burban:coh0} \sum_{n=0}^{\infty}\frac{(-1)^n
q^{-n(n-1)/2}}{(q;q)_n}h^{(0)}_n(x;q)t^n = (q
t;1/q)_{\infty}\,{}_{1}\phi_1\Bigl(\begin{array}{c|c}x\\t
q\end{array}\,1/q; -tq\Bigr).
\end{equation}
Let us denote the left hand side of this identity by
\begin{equation}\Phi (x,q,t) = \sum_{n=0}^{\infty}\frac{(-1)^n
q^{-n(n-1)/2}}{(q;q)_n}h^{(0)}_n(x,q)t^n.
\end{equation}
Then \begin{equation} \Phi({\rm i}x,1/q,t)= \sum_ {n =
0}^{\infty}\frac{(-1)^n q^{n(n-1)}}{(q;q)_n}{\tilde
h}^{(0)}(x;q)(\frac{it}{q})^n
\end{equation}
Using formula (3.29.12) of ref. ~\cite{KS}
\begin{equation}
\sum_{n =0}^{\infty}\frac{(-1)^n q^{n(n-1)}}{(q;q)_n}{\tilde
h}^{(0)}(x;q)t^n = (-{\rm i}t;q)_{\infty}\,{}_1\phi_1 \Bigl(
\begin{array}{c|c}{\rm i}x\\-{\rm i}t
\end{array}\,
q;{\rm i}t\Bigr)
\end{equation}
we obtain
\begin{equation}\Phi({\rm i}x,1/q,t)=
(-t/q;q)_{\infty}\,{}_1\phi_1 \Bigl(
\begin{array}{c|c}{\rm i}x\\t/q
\end{array}\,q;-t/q \Bigr)
\end{equation}
from which easy follows (\ref{burban:coh0}).

The Barut-Girardello coherent states of a oscillator
(\ref{burban:cond0}), (\ref{burban:cond1}) in the Fock
representation space ${\cal H}$ are defined as eigenvectors of
annihilation operator $a:$
\begin{equation}
\label{burban:def1} a\,|z\rangle = z\,|z\rangle,\quad z\in
\mathbb{C},
\end{equation} given by the formula
\begin{equation}
\label{burban:coh} |z\rangle = {\cal
N}^{-1}\sum_{n=0}^{\infty}\frac {z^n}{f_ {n-1}!}|n\rangle,
\end{equation} where
${\cal N}$ is normalized factor. By definition the basis vectors
$|n\rangle $ of ${\cal H}$ are taken as polynomials
$\psi^{(0)}_n(x;q)$ of (\ref{burban:basis0}).
\medskip
In the case of the oscillator corresponding q-Hermite I
polynomials we have $f_{n-1} = \sqrt{1/(1-q))q^{n-1}(1-q^n)}$. It
follows
\begin{equation}
f_{n-1}! = \sqrt{1/(1-q)^n q^{n(n-1)/2}(q;q)_n}
\end{equation} and coherent state (\ref{burban:coh}) can be written as
\begin{equation}
|z\rangle = {\cal N}^{-1}(|z|^2)\sum_{n =
0}^{\infty}\frac{(\sqrt{1-q}z)^n
}{\sqrt{q^{n(n-1)/2}(q;q)_n}}\frac{q^{-n(n-1)/4}}{(q;q)_n^{1/2}}h^{(0)}_n(x;q)
\end{equation}
\begin{equation}
= {\cal N}^{-1}(|z|^2)\sum_{n = 0}^{\infty}\frac{(-1)^n
q^{-n(n-1)/2}}{(q;q)_n}h^{(0)}_n(x,q)(-\sqrt{1-q}z)^n
\end{equation}
(take into account of (\ref{burban:coh0}))
\begin{equation}
= {\cal N}^{-1}(|z|^2)(q
(-\sqrt{1-q}z);1/q)_{\infty}\,{}_{1}\phi_1\Bigl(\begin{array}{c|c}x\\(-\sqrt{1-q}z)
q\end{array}\,1/q;q\sqrt{1-q}z \Bigr).
\end{equation}
Easy calculation gives the normalized factor
\begin{equation}
{\cal N}^{2}(|z|^2)= {}_2\phi_0 \Bigl( \begin{array}{cc|c}
0,&0&\\{} &-&\end{array}\,q;(1-q)z\Bigr).
\end{equation}
The overlapping of two coherent states is
\begin{equation}\langle z_1|z_2\rangle =
{}_2\phi_0 \Bigl( \begin{array}{cc|c} 0,&0&\\{}
&-&\end{array}\,q;(1-q){\bar z}_1z_2\Bigr).
\end{equation}

The terminal expression for coherent state $|z\rangle$ of the
q-oscillator corresponding discrete q-Hermite I polynomials has
the form $$|z\rangle = \Bigl\{{}_2\phi_0 \Bigl(
\begin{array}{cc|c} 0,&0&\\{}
&-&\end{array}\,q;(1-q)z\Bigr)\Bigr\}^{1/2} $$
\begin{equation}
\times(-\sqrt{1-q}z);1/q)_{\infty}\,{}_{1}\phi_1\Bigl(\begin{array}{c|c}x\\
-q\sqrt{1-q}z\end{array}\,1/q;q\sqrt{1-q}z \Bigr).
\end{equation}

The family of coherent states associated with oscillator-like
system corresponding to generalized discrete q-Hermite I
polynomials (\ref{burban:finitaI}) can be obtained the same
method. In this case
\begin{equation}
f_ {n-1}! = \sqrt{1/(1-q')^n q^{\alpha n^2} q^{(\alpha +
\beta)n}(q';q')_n}
\end{equation}
and basis vectors $|n\rangle$ are taken as polynomials
$\psi_n(x;q)$ of (\ref{burban:basis1}). This leads
(\ref{burban:coh}) to
$$|z\rangle = {\cal N}^{-1}(|z|^2)\sum_{n =
0}^{\infty} \frac{(\sqrt{1-q'}z)^n}{\sqrt{ q^{\alpha n^2}
q^{(\alpha + \beta)n}(q';q')_n}}\,\frac {q^{-\alpha
n^2/2}h_n(x;q)}{q^{(\alpha + \beta) n/2}(q';q')_n^{1/2}},$$ or
\begin{equation}
\label{burban:chr} |z\rangle = {\cal N}^{-1}(|z|^2)\sum_{n =
0}^{\infty} \frac {(-1)^n q^{-\alpha
n(n-1)}}{(q;q)_n}h_n(x;q)\Bigl( -\frac{\sqrt{1-q'}}{q^{2\alpha +
\beta}}z \Bigr)^n
\end{equation}

The generating function for generalized discrete q-Hermite I
polynomials (the extension of (\ref{burban:coh0})) is given by the
formula $$ \sum_{n=0}^{\infty}\frac{(-1)^n q^{-(2\alpha + \beta)n}
q^{-\alpha n(n-1)}}{(q;q)_n}h_n(x,q)t^n $$
\begin{equation}
\label{burban:gcohr0} = (t
q';1/q')_{\infty}\,{}_{1}\phi_1\Bigl(\begin{array}{c|c}q^{-(2\alpha
n + \beta)/2}q'^{(n-1)/2} x\\q't
\end{array}\,1/q';-tq'\Bigr).
\end{equation}
Comparing the expressions (\ref{burban:chr}) and
(\ref{burban:gcohr0}) we obtain $$|z\rangle = {\cal
N}^{-1}(|z|^2)(-q^{-(2\alpha+\beta)}q'\sqrt{1-q'}z;1/q')_{\infty}
$$
\begin{equation}
\label{burban:gcoh1} \times {}_{1}\phi_1\Bigl(\begin{array}{c|c}
q^{-(2\alpha n + \beta)/2}q'^{(n-1)/2} x\\- q^{-(2\alpha +
\beta)}q'\sqrt{1-q'}z
\end{array}\,1/q';q^{-2\alpha + \beta}q'\sqrt{1-q'}z\Bigr).
\end{equation}
The short calculation of a normalizing factor of the coherent
state (\ref {burban:gcoh1}) gives
\begin{equation}
{\cal N}^2(|z|^2) =
\sum_{n=0}^{\infty}\Bigl(\frac{\sqrt{1-q'}}{q^{2\alpha +
\beta}}|z|^2\Bigr)^n\frac{q^{-\alpha n(n-1)}}{(q';q')_n}.
\end{equation}
The description of the Barut-Girardello type coherent states for
the oscillator-like systems, connected with discrete q-Hermite II
polynomials has been done in \cite{BD}. Therefore in the further
we consider the coherent states, connected only with
oscillator-like systems connected with generalized discrete
q-Hermite II. They are given by the expression (\ref{burban:coh}),
where
\begin{equation}
f_{n-1}! = \sqrt{q^{\beta n}/(1-q')^n q^{\alpha n(n +
1)}(q';q')_n}
\end{equation} and basis vectors $ |n\rangle $ are given by   ${\tilde
\psi}_n(x,q)$ of (\ref{burban:basis2}), namely,
\begin{equation}
|z\rangle = {\cal N}^{-1}\frac {z^n}{\sqrt{q^{\beta n}/(1-q')^n
q^{\alpha n(n + 1)}(q';q')_n}}\,\frac{q^{-\alpha
n^2/2}}{q^{(2\alpha + \beta)n/4)}(q';q')_n^{1/2}}{\tilde
h}_n(x;q),
\end{equation}
or
\begin{equation}\label{burban:gcoh5}
|z\rangle = {\cal N}^{-1}(|z|^2)\sum_{n = 0}^{\infty} \frac
{(-1)^n q^{-\alpha n^2}}{(q;q)_n}h_n(x;q)\Bigl(- q^{-\beta/4}
q^{-(2\alpha + \beta)/2} \sqrt{1-q'}z \Bigr)^n.
\end{equation}
The generation function for polynomials ${\tilde h_n(x;q)}$ is
defined by  $$ \sum_{n = 0}^{\infty}\frac{(-1)^n q^{-\alpha
n^2}}{(q;q)_n}h_n(x;q)\,t^n $$
\begin{equation}
\label{burban:gcoh3} = (-{\rm i} q^{\beta/4} q'^{1/2}t;
{q'})_{\infty}\,{}_{1}\phi_1\Bigl(\begin{array}{c|c}{\rm
i}q^{-(2\alpha n + \beta)/2}q'^{(2n-1)/2} x\\-{\rm
i}q^{\beta/4}q'^{\beta/4}t
\end{array}\,{q'};{\rm i}tq^{\beta/4}q'^{1/2}\Bigr)
\end{equation} (the extension of the formula (3.29.12) of ref.
\cite{KS}). Using this identity the coherent state $|z\rangle$ of
(\ref{burban:gcoh5}) can represented by expression $$|z\rangle
={\cal N}^{-1}(|z|^2) $$
\begin{equation}
\label{burban:gcoh3}\times({\rm
i}\sqrt{\frac{q'(1-q')}{q^{(2\alpha + \beta)/2}}}z;
{q'})_{\infty}\, {}_{1}\phi_1\Bigl(\begin{array}{c|c}{\rm
i}q^{-(2\alpha n + \beta)/2}q'^{(2n-1)/2} x\\{\rm
i}\sqrt{\frac{q'(1-q')}{q^{(2\alpha + \beta)/2}}}z
\end{array}\,{q'};{-\rm i}\sqrt{\frac{q'(1-q')}{q^{(2\alpha + \beta)/2}}}z\Bigr).
\end{equation}

At last, the normalizing factor of the generalized coherent state
(\ref{burban:gcoh3}) can be written as
\begin{equation}
{\cal N}^2(|z|^2) = \sum_{n =0}^{\infty}\frac{q^{-\alpha
n(n+1)}}{(q';q')_n}\Bigl(\frac{1-q'}{q^{\beta}}|z|^2\Bigr)^n.
\end{equation}

We have not established completeness (over-completeness) of the
given set of generalized coherent states. It will be done in a
forthcoming paper.
\medskip
\medskip
\medskip

\subsection*{Acknowledgements}
I would like to thank A.U. Klimyk for many useful discussions,
valuable suggestions and observations. This research was partially
supported by Grant 10.01/015 of the State Foundation of
Fundamental Research of Ukraine.

\end{document}